\documentclass[final,3p,times,twocolumn]{elsarticle}

\usepackage{amssymb}
\usepackage{bm}

\journal{Non-crystalline Solids}

\usepackage{color}

\begin{document}

\begin{frontmatter}



\title{Glass transition of binary mixtures of dipolar particles in two dimensions}


\author{David Hajnal}
\author{Martin Oettel}
\author{Rolf Schilling}
\address{Institut f\"ur Physik, Johannes Gutenberg-Universit\"at Mainz, Staudinger Weg 7, D-55099 Mainz, Germany}

\begin{abstract}
We study the glass transition of binary mixtures of dipolar particles
in two dimensions within the framework of mode-coupling theory, focusing in particular on
the influence of composition changes. In a first step, we demonstrate that the experimental system
of K\"onig et al. [Eur. Phys. J. E {\bf 18}, 287 (2005)] is well described by point dipoles through a comparison
between the 
experimental partial structure factors and those from our Monte Carlo simulation.
For such a mixture of point particles we show that there is 
always a plasticization effect, i.e. a stabilization of the liquid state  due to mixing, in contrast to binary hard disks.
We demonstrate that the predicted plasticization effect is in qualitative agreement with experimental results.
Furthermore, also some general
properties of the glass transition lines are discussed.

\end{abstract}


\end{frontmatter}


\section{Introduction}

The study of supercooled liquids close to the glass transition is
a rapidly developing field in modern physics. For both experimental and
theoretical investigations, colloidal dispersions are often chosen as model systems.
Compared to atomic liquids, the main advantages of such model systems rely
on their simplicity: the particle interactions can often be described theoretically
by quite simple and well defined expressions, and experimental investigations
can be performed by applying standard optical methods in real space  
instead of neutron or X-ray scattering techniques.

Supercooled liquids close to the glass transition exhibit complex relaxation patterns 
which do not follow a simple exponential decay in time.
The most successful theory for the description of the dynamics of 
supercooled liquids is the so-called mode-coupling theory (MCT) which has been developed and studied
in  great detail by G\"otze and coworkers \cite{Goetze09} for various model systems
in three spatial dimensions (3D). The central prediction of MCT is a transition from
a liquid into an ideal glassy state upon decreasing temperature or increasing particle density
below or above some critical value, respectively.

It is well-known that many physical phenomena like equilibrium phase transitions depend
strongly on the spatial dimensionality $d$.
Consequently, it appears to be essential to study the $d$-dependence of the liquid-glass transition as well.
As far we know, the first experimental approach to this issue is the investigation of K\"onig et al. \cite{Koenig05}
of a two-dimensional (2D)
binary mixture
of super-paramagnetic colloidal particles which are trapped on a water-air interface. They interact 
via repulsive dipole potentials. Here, the magnetic dipoles have been induced by an external magnetic 
field perpendicular to the water interface which renders the repulsions isotropic in the interface plane.  
Their results for the
self-intermediate scattering functions measured by video microscopy clearly exhibit the stretched relaxation
patterns of glass-forming liquids and the corresponding strong slowing down of the dynamics 
when increasing the magnetic field (which plays the role of an inverse, effective temperature).
This model system was also used for both experimental and theoretical studies on
crystallization effects in the high-coupling limit, i.e. for large magnetic fields (see \cite{Assoud09} and references therein).
A further interesting phenomenon (found experimentally and
corroborated theoretically) pertaining to the same model system in the liquid state
is a partial clustering of the smaller particles in a sponge-like topology  \cite{Hoffmann06}.

The first MCT calculations on a model system in 2D were performed by Bayer et al. \cite{Bayer07}
for monodisperse hard disks.
An ideal glass transition has been found at a critical packing fraction $\varphi^c$.
This study was extended to binary mixtures of hard disks \cite{Hajnal09}
characterized by the total packing fraction $\varphi$, the concentration of the smaller disks $x_s$, and the size ratio $\delta$.
The same four mixing effects  occur as have been
reported before by G\"otze and Voigtmann \cite{Voigtmann03} for binary mixtures of hard spheres in 3D: 
{(i) for small size disparities the glass is stabilized, (ii) for large size disparities the liquid state is stabilized,
(iii) upon increasing the concentration of the smaller particles
the plateau values of the normalized correlation functions for intermediate times increase
for almost all wave numbers accompanied by (iv) a slowing down of the initial part of the
relaxation of the correlators of the bigger particles towards these plateaus.}
Further, it was shown that the glass transition diagram for binary hard disks, i.e. the critical packing fraction $\varphi^c$
as function of $x_s$ for several $\delta$,
strongly resembles the
corresponding random close packing diagram. This fact is an indication for the validity of 
the MCT approximations also in 2D.

In the present study, we will apply MCT to a 2D binary liquid of dipoles. Our aim is twofold. First, we want to explore how 
far the mixing effects found for hard spheres \cite{Voigtmann03} and hard disks \cite{Hajnal09} have some
generic features. Second, motivated by the experiments on
quasi-2D binary mixtures of super-paramagnetic particles \cite{Koenig05},
we want to compare our MCT results with available data. 
As a byproduct, we propose a simple empirical
ansatz for the bridge functions \cite{hansen} appearing in the closure to the Ornstein-Zernike (OZ) equation
which allows for a rather precise determination of the equilibrium structure.

The organization of our contribution is as follows. Sect.~\ref{theory} will contain (i) a justification of the model
used to approach the experimental system, based on a comparison of the experimental and theoretical results for the static
structure factors, (ii) a concise review of MCT equations in $d$ dimensions, and (iii) the discussion of some properties
of the glass transition lines.
Our results from MCT will be presented and discussed in Sect.~\ref{results} and the final Sect.~\ref{conclusions}
summarizes and gives some conclusions.

\section{Model system and mode-coupling theory}

\label{theory}

In this section, we will introduce the model being investigated, discuss some of its structural properties,
and shortly review the corresponding MCT equations. For more details on the latter
the reader may consult Ref.~\cite{Hajnal09}.

Let us summarize some basic notation.
We make use of the compact mathematical notation introduced in Ref.~\cite{Hajnal09}.
Bold symbols $\boldsymbol{A}$, $\boldsymbol{B}$ etc.
{denote arrays of matrices}. Their components $\boldsymbol{A}_k$, $\boldsymbol{B}_k$
labeled by subscript Latin indices are $M\times M$ matrices.
Their elements $A_k^{\alpha\beta}$, $B_k^{\alpha\beta}$,
also denoted by $(\boldsymbol{A})_k^{\alpha\beta}$, $(\boldsymbol{B})_k^{\alpha\beta}$ when appropriate,
are indicated by superscript Greek indices.
Matrix products are defined component-wise, e.g. $\boldsymbol{C}=\boldsymbol{A}\boldsymbol{B}$ means $\boldsymbol{C}_k=\boldsymbol{A}_k\boldsymbol{B}_k$ for all $k$.
$\boldsymbol{A}$ is called positive (semi-)definite,
($\boldsymbol{A}\succeq\boldsymbol{0}$) $\boldsymbol{A}\succ\boldsymbol{0}$ if this is true for all $\boldsymbol{A}_k$.

\subsection{The model system}

Our starting point is the experimental model system studied by K\"onig et al. \cite{Koenig05}.
Technical details for the experimental setup can be found in Ref.~\cite{Ebert09}.
The authors consider a binary mixture of super-paramagnetic colloidal particles trapped on a water-air interface
of a pending droplet,
so that particle motion is effectively only possible in 2D. By applying an external magnetic field perpendicular
to the liquid interface, they can vary the strength of the repulsive dipolar interaction forces between
the particles. Temperature $T$, total particle number density $n$, radii $R_{\alpha}$ and susceptibilities $\chi_{\alpha}$
of the smaller ($\alpha=s$) and bigger ($\alpha=b$) particles are kept constant to good accuracy.
To observe vitrification, the magnetic field is increased.
Accordingly, the magnetic moments of the particles and the repulsions between them 
grow and in turn the correlations between the particles become stronger.

The colloidal dispersion is strongly diluted. To be more precise, the average distance between the particles
exceeds the diameter of the bigger particles roughly by a factor of four \cite{Koenig05}, which yields the estimate
$\varphi<(4\cdot2R_b)^{-2}\pi R_b^2=\pi/64\approx0.05$ for the total 2D-packing fraction on the interface plane.
Therefore, collisions of the hard cores of two particles practically never occur, at least
for low enough temperatures.
Under this condition, the thermodynamic equilibrium state of the experimental system is well determined
by the induced dipole potentials, only.

Idealizing this picture,
we assume a binary mixture of point particles in 2D interacting via the induced dipole potentials
\begin{equation}
\label{potential}
u^{\alpha\beta}(r)=\frac{\mu_0}{4\pi}\frac{\chi_{\alpha}\chi_{\beta}B^2}{r^3}.
\end{equation}
$\mu_0$ is the vacuum permeability and
$B$ denotes the applied external magnetic field.
Although the hard core interactions have been neglected, the two species of particles shall be called ``big'' ($\alpha=b$) and ``small'' ($\alpha=s$)
in the sense that $\chi_{b}\geq\chi_{s}$.

The thermodynamic state of the considered system with fixed $\chi_b$ and $\chi_s$ depends on four variables:
$n$, $T$, the concentration of the smaller
dipoles $x_s$, and the magnetic field $B$.
For a monodisperse dipolar system with susceptibility $\chi_0$ it is well-known that the dependence on $T$, $n$,
$B$ is given by the dimensionless coupling parameter
\begin{equation}
\label{gamma_mono}
 \Gamma_0=n^{3/2}\frac{\mu_0}{4\pi}\frac{\chi_0^2B^2}{k_BT}.
\end{equation}
Since we want to study mixing effects on the ideal glass transition, we have to compare e.g. the critical
temperature $T^c(x_s)$ or equivalently the critical filed $B^c(x_s)$ for the binary system with concentration
$0<x_s<1$ with a corresponding monodisperse system. {As we will argue now it does not make sense  for all $x_s$
to simply choose for that comparison the monodisperse system with susceptibility $\chi_b$ or $\chi_s$.}

{The magnetization per site $m$ is a relevant physical quantity. For the binary system it is
$m_{binary}(x_s)=[x_s\chi_s+(1-x_s)\chi_b]B$. Now we will consider a monodisperse system with a susceptibility
$\chi(x_s)$ for \textit{all} point dipoles. Then it is $m_{mono}(x_s)=\chi(x_s)B$.
For a comparison of the binary and the monodisperse system it is reasonable to require that both have
the same magnetization. From this condition we get $\chi(x_s)=x_s\chi_s+(1-x_s)\chi_b$ and finally the
generalization of Eq.~(\ref{gamma_mono}) to binary dipolar systems in 2D:}
\begin{equation}
\label{Gamma}
\Gamma=(\pi n)^{3/2}\frac{\mu_0}{4\pi}\frac{[x_s\chi_s+(1-x_s)\chi_b]^2B^2}{k_BT},
\end{equation}
which is the dimensionless interaction parameter chosen by K\"onig et al \cite{Koenig05}. In the following
we will use this parameter, 
the susceptibility ratio $\delta=\chi_s/\chi_b\leq1$, and the concentration $x_s$ of the smaller dipoles
as independent control parameters. The prefactor $\pi^{3/2}$ in Eq.~(\ref{Gamma}) was
introduced to match the convention used in Ref.~\cite{Koenig05}.

\subsection{Static structure functions}

The interaction potentials for the specified model system enter the MCT equations only via the static structure factors.
Thus, we need some accurate and efficient method for a systematic calculation of these objects. We have at least two possibilities.

First, we can make use of the well established integral equation theory for the direct correlation functions
$c_k^{\alpha\beta}$ based on the OZ equation \cite{hansen} for mixtures 
\begin{equation}
\boldsymbol{h} = \boldsymbol{c}+n\boldsymbol{c}\boldsymbol{x}\boldsymbol{h}
\label{oz}
\end{equation}
where $x_k^{\alpha\beta}=x_{\alpha}\delta_{\alpha\beta}$ and $h^{\alpha\beta}_k$ are the total correlation functions
for given wave number $k$.
$\delta_{\alpha\beta}$ denotes the Kronecker delta.
To obtain e.g. $\boldsymbol{c}$, we need a closure relation. The general {ansatz \cite{hansen} for this} reads
\begin{equation}
\label{closure}
\ln[g^{\alpha\beta}(r)]=-u_{eff}^{\alpha\beta}(r)+h^{\alpha\beta}(r)-c^{\alpha\beta}(r),
\end{equation}
\begin{equation}
\label{bridge}
u_{eff}^{\alpha\beta}(r)=\frac{u^{\alpha\beta}(r)}{k_BT}-d_0^{\alpha\beta}(r).
\end{equation}
$g^{\alpha\beta}(r)=h^{\alpha\beta}(r)+1$ are the pair distribution functions and $d_0^{\alpha\beta}(r)$ are called
the (still unknown) bridge functions. Every specific closure relation relies on some (uncontrolled)
approximation scheme for $d_0^{\alpha\beta}(r)$. The advantage of using an integral equation theory is, however,
that data with high numerical quality can be produced. There is no statistic noise and there are
no finite size effects like in atomistic computer simulations. Furthermore, crystallization effects can explicitly
be excluded. Having found $\boldsymbol{c}$, the static correlation matrix $\boldsymbol{S}$ follows from
$
(\boldsymbol{S}^{-1})_k^{\alpha\beta}=\delta_{\alpha\beta}/x_{\alpha}-nc_k^{\alpha\beta}.
$

Second, we can perform an atomistic computer simulation. Since we are only interested in static quantities,
the most efficient choice is the Monte Carlo (MC) technique. It yields exact numerical results in the sense that
these do not rely on uncontrolled approximations. However, here are also some limitations.
In order to study glassy behavior,
we have to restrict the
range of the control parameters such that no {crystallization effects} occur. Furthermore, we need data with high numerical
quality as input for MCT. For instance, the violation of the strict positive definiteness of the static
correlation matrix $\boldsymbol{S}_k$
for a single $k$ already leads to instabilities in the numerical solution of the MCT equations. Such problems typically
occur only for the atomistic computer simulation data at 
small $k$ where both statistical noise and finite size effects may become dominant.
Thus, we need a clearly defined way to smooth the numerical raw data such that $\boldsymbol{S}\succ\boldsymbol{0}$ is enforced.

\begin{figure}
\includegraphics[width=1\columnwidth]{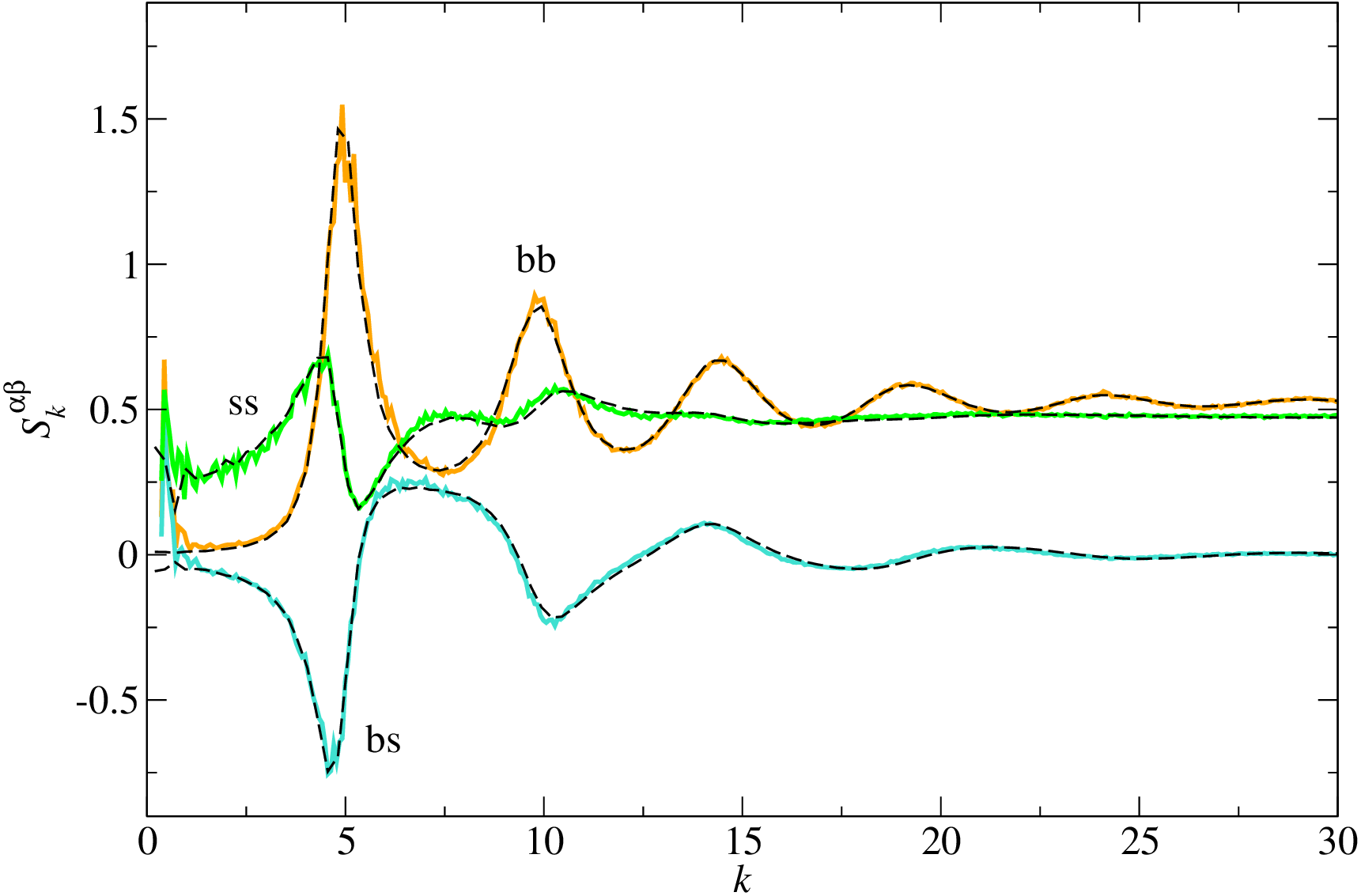}
\caption{(Color online) Partial structure factors
for a binary mixture of dipolar particles in 2D at $\Gamma=110$, $\delta\approx0.1$, and $x_s=0.48$
measured experimentally by Ebert \cite{ebert_phd}
{(solid lines) and the corresponding MC results (dashed lines)} obtained with $600$ particles without data
smoothing.
}
\label{fig_mc}
\end{figure}

Now, let us come back to our specific model system.
{We are able to reproduce available experimental data
for $\boldsymbol{S}_k$ by performing MC simulations such that the deviations are smaller
than the noise level of the  experimental data.}
Fig.~\ref{fig_mc} shows partial structure factors $S_k^{\alpha\beta}$
for a binary mixture of dipolar particles in 2D at
$\Gamma=110$, $\delta\approx0.1$, and $x_s=0.48$
measured experimentally by Ebert \cite{ebert_phd}.
Due to the finite polydispersity of both species of colloidal particles the
value for $\delta$ is approximative, only.
Also included are corresponding MC results which we have obtained
as described in \ref{mc} with two modifications: First, we have produced these MC data by using only
$N=600$ particles instead of $N=1600$, and second, we have not smoothened the raw data. Thus, we observe that for low $k$
(especially for $S_k^{ss}$) the MC structure factors are superimposed by oscillations stemming from an interplay of
uncertainties of $h^{\alpha\beta}(r)$ at large $r$ and the real space cutoff $R$ for its Fourier transform.
Besides these technical artifacts, our MC results are in very good agreement with the experimental data.
Note that there are no adjustable parameters.
For $N=600$ particles, we were also able to reproduce experimentally measured partial structure factors within the parameter range
$\delta\approx0.1$,
$0.29\leq x_s\leq 0.48$,
and
$51\leq\Gamma\leq281$
which are not shown here.
At smaller $\Gamma$, the influence of the hard-core interactions of the particles seems to become significant in the experimental data.
For larger $\Gamma$, we would need larger system sizes and longer simulation runs in order to reproduce the experimental data.
We can draw two important conclusions. First, for $\Gamma>51$ the experimental system of K\"onig et al. \cite{Koenig05}
is very well described by the idealized model potentials given by Eq.~(\ref{potential}).
Second, we can be confident that our MC simulations with $1600$ particles will yield sufficiently realistic results
for our purposes, since the MCT glass transitions discussed below occur within the parameter range $80<\Gamma<160$.

Using integral equation theory, we have further analyzed several established choices for the
bridge functions $d_0^{\alpha\beta}(r)$ in Eq.~(\ref{bridge}). The Percus-Yevick (PY) ansatz, which has successfully
been applied to
systems with hard core interactions,
{fails in describing MC simulation data for our system.
For instance, for a binary mixture with $\delta=0.1$, $x_s=0.5$, and $\Gamma=10$, i.e. deep in the liquid regime,
the PY closure underestimates the position of the main peak of $S_k^{bb}$ by about $10$ percent and overestimates the
maximum value of this peak by almost $100$ percent.
For higher values of $\Gamma$ we have not found stable numerical solutions for the PY structure factors.
Accordingly we do not know whether the quality of the PY theory for our model system becomes better or even worse.
At least for $\Gamma\leq10$, PY theory is much less
reliable for dipolar potentials than for hard core interactions. The reason for this is not obvious.}
Another choice is the hypernetted chain (HNC)
approximation $d_0^{\alpha\beta}(r)=0$.
This simple ansatz describes the positions of the main peaks of the partial
structure factors correctly, but strongly underestimates their amplitudes (see below),
as already noticed by 
Zahn \cite{zahn_phd} and Hoffmann \cite{hoffmann_phd}.
There are also more sophisticated closure relations. Such relations, however, typically include free adjustable parameters.
This fact makes these methods not only computationally demanding, but also makes the 
optimization procedure unwieldy
if more than
one adjustable parameter is used. The Rogers-Young (RY) closure, for instance, {which interpolates}
between PY and HNC,
has in principle three fit parameters for the binary mixture. Furthermore, the quality of the corresponding results is not
satisfactory. Fig.~3(a) in Ref.~\cite{Hoffmann06} clearly shows that the RY closure underestimates the position of the main
peak of the partial structure factors of the big dipoles. It also underestimates the amplitude of the oscillations
beyond the main peak.

For our binary dipole model in 2D, we have found empirically that using HNC with temperature $T/2$ instead of $T$ leads
to a systematic improvement in the description of experimental and simulated data for $\boldsymbol{S}_k$
(see below). A similar observation was made earlier independently by Zahn \cite{zahn_phd}.
Even more, it was observed by Klapp \cite{klapp_unpublished} that also simulation results for specific dipolar
systems in 3D can be well fitted by using HNC with $T/2$. These observations let us
propose a simple empirical ansatz for the bridge functions:
\begin{equation}
\label{bridge2}
d_0^{\alpha\beta}(r)=-\frac{u^{\alpha\beta}(r)}{k_BT}.
\end{equation}
Since Eq.~(\ref{bridge2}) has no adjustable parameters, it is well suited for fast and systematic calculations.
Eq.~(\ref{bridge2}) will be called the $T/2$-HNC closure.

In the following, we will use both, MC and $T/2$-HNC, for calculating the static structure input for MCT.
Since the MC results for $S_k^{\alpha\beta}$ fit the corresponding experimental ones very well,
this ensures that the observed effects are {predictions of MCT} and not just artifacts
of one of the specific techniques for the calculation of the static input.
To calculate $T/2$-HNC structure factors numerically, we first eliminate the total correlations functions in Eqs.~(\ref{oz}) and (\ref{closure})
by introducing the new function $\boldsymbol{\gamma}=\boldsymbol{h}-\boldsymbol{c}$ and apply then the Lado algorithm \cite{Lado}
with a real space cutoff $R\sqrt n =100$  and $N_{grid}=4000$ grid points.
The MC simulations are performed in the canonical ensemble with, if not explicitly stated otherwise, $N=1600$ particles and $8$ image boxes.
Technical details and the procedure to smooth the raw data are explained in \ref{mc}.

\begin{figure}
\includegraphics[width=1\columnwidth]{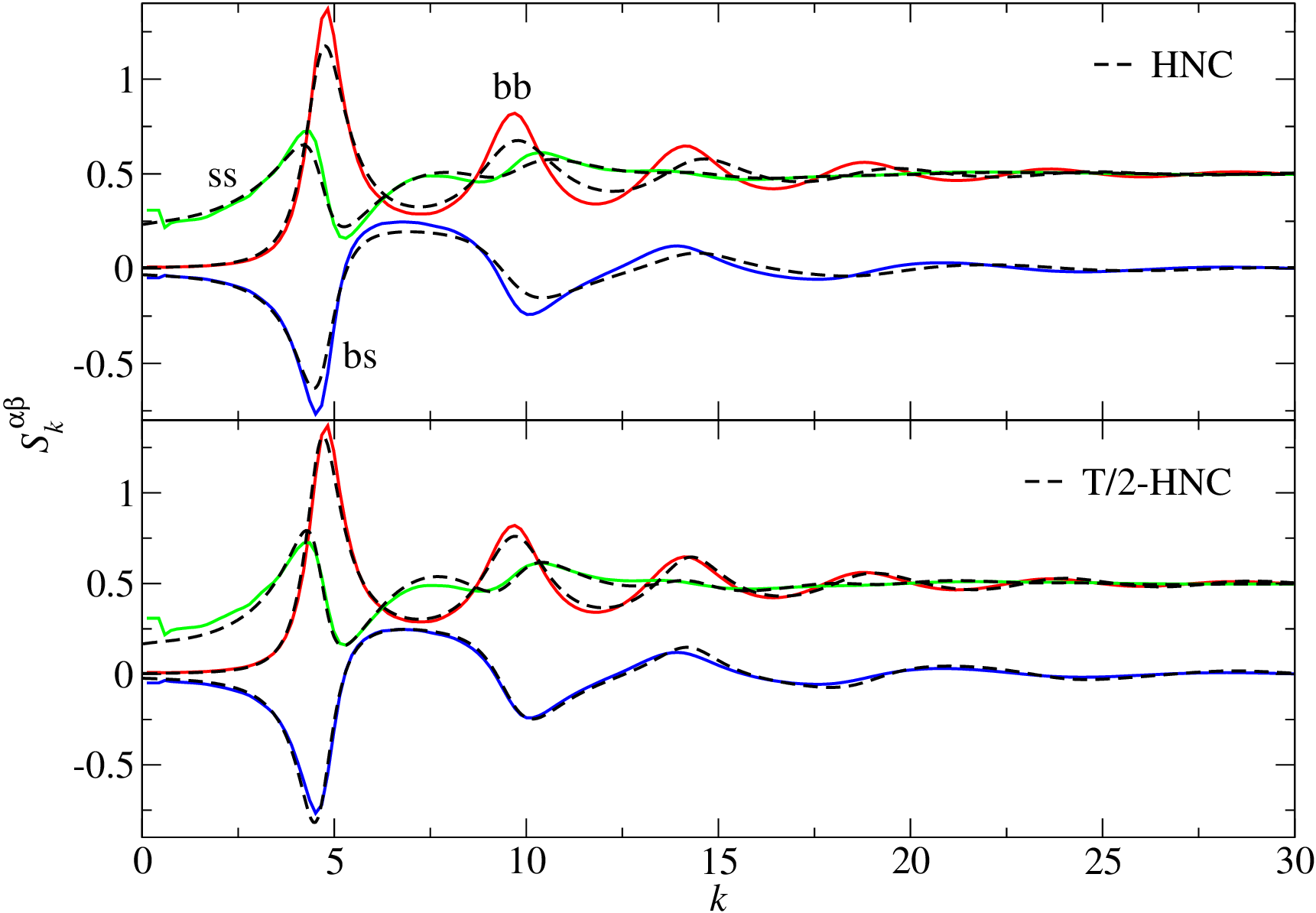}
\caption{(Color online) Partial structure factors
for a binary mixture of dipolar particles in 2D
at $\Gamma=120$, $\delta=0.1$, and $x_s=0.5$ from MC simulation with $1600$ particles
(solid lines) and for HNC (dashed lines in upper panel) and $T/2$-HNC (dashed lines in lower panel).
}
\label{fig_hnc}
\end{figure}

Fig.~\ref{fig_hnc} shows MC simulated partial structure factors for the binary dipole model in 2D at $\Gamma=120$, $\delta=0.1$, and $x_s=0.5$
compared to corresponding HNC and $T/2$-HNC results. While the HNC closure underestimates the
amplitude of the oscillations of the partial structure factors (see upper panel of Fig.~\ref{fig_hnc}), the $T/2$-HNC results are in a surprisingly good
agreement with the MC data (see lower panel of Fig.~\ref{fig_hnc}). The agreement is not perfect since there are some smaller deviations, but
the qualitative overall trends are even better than for the RY closure. Remember that the RY closure underestimates
the position of the main peak of the partial structure factors of the big dipoles and also underestimates the amplitude of the oscillations
beyond the main peak (see Fig.~3(a) in Ref.~\cite{Hoffmann06}). The $T/2$-HNC data in Fig.~\ref{fig_hnc}
do not exhibit such qualitative drawbacks.
Furthermore, we have found
for all investigated examples within the control parameter range
$0.1\leq\delta\leq0.6$,
$0.1\leq x_s\leq 0.9$,
and
$20\leq\Gamma\leq 160$
that, compared to the standard HNC closure, the $T/2$-HNC closure leads to a similar improvement in the description of
MC data  as shown in Fig.~\ref{fig_hnc}.
In the next section, this trend will also be strongly supported by
the fact that using MC and $T/2$-HNC structure factors as input for MCT will lead to compatible
glass transition scenarios on very similar temperature scales.

The MC data in Fig.~\ref{fig_hnc} were obtained by simulating $N=1600$ particles and
smoothing the raw data as described in \ref{mc}. The positive definiteness $\boldsymbol{S}\succ\boldsymbol{0}$ is
guaranteed. However, there remain still some artifacts at the lowest $k$ values,
see especially the kinks in $S_k^{ss}$ for $k<0.8$.
By testing different extrapolation methods we have found that these artifacts do not influence the glass transition scenarios shown below,
and hence can be ignored.

\subsection{MCT equations}

The object of main interest for a statistical description of
an isotropic and homogeneous classical fluid with $M$ macroscopic components
is the matrix  $\boldsymbol{\Phi}(t)$
of time-dependent partial
autocorrelation functions of density fluctuations, 
$\Phi_k^{\alpha\beta}(t)$ ($\alpha,\beta=1,\dots,M$), at wave number $k$.
The relation to static structure is given by $\boldsymbol{\Phi}(0)=\boldsymbol{S}$, where 
the elements of the static structure factor matrix $\boldsymbol{S}$
obey
$\lim_{k\rightarrow\infty}S_{k}^{\alpha\beta}=x_{\alpha}\delta_{\alpha\beta}$.

Starting point for the calculation of $\boldsymbol{\Phi}(t)$ is the exact Zwanzig-Mori equation.
Assuming overdamped colloidal dynamics, it reads 
\begin{equation}
\label{zwanzig_mori}
\boldsymbol{\tau}\boldsymbol{\dot{\Phi}}(t)+\boldsymbol{S}^{-1}\boldsymbol{\Phi}(t)+\int_0^t\mathrm{d} t'
\boldsymbol{m}(t-t')\boldsymbol{\dot{\Phi}}(t')=\boldsymbol{0}.
\end{equation}
The positive definite matrix $\boldsymbol{\tau}$ of microscopic relaxation times will not be of further interest
in this paper.
MCT approximates $\boldsymbol{m}(t)$ by a symmetric bilinear functional
\begin{equation}
\boldsymbol{m}(t)=\boldsymbol{\mathcal{F}}[\boldsymbol{\Phi}(t),\boldsymbol{\Phi}(t)].
\end{equation}
For a multicomponent liquid in $d\geq2$ dimensions it reads explicitly \cite{Hajnal09}
\begin{eqnarray}
\label{memory}
\mathcal{F}_k^{\alpha\beta}[\boldsymbol{X},\boldsymbol{Y}]&=&\frac{\Omega_{d-1}}{(4\pi)^d}
\sum_{\alpha',\beta',\alpha'',\beta''}\int_0^\infty \mathrm{d} p\int_{|k-p|}^{k+p} \mathrm{d} q\nonumber\\
&&\times V^{\alpha\beta;\alpha'\beta',\alpha''\beta''}_{k;p,q}X_p^{\alpha'\beta'}Y_q^{\alpha''\beta''}
\label{bilinear}
\end{eqnarray}
where the vertices are given by
\begin{equation}
\label{vertex_1}
V^{\alpha\beta;\alpha'\beta',\alpha''\beta''}_{k;p,q}=
\frac{n}{x_{\alpha}x_{\beta}}\frac{pq}{k^{d+2}}v_{kpq}^{\alpha\alpha'\alpha''}v_{kpq}^{\beta\beta'\beta''},
\end{equation}
\begin{eqnarray}
\label{vertex}
v^{\alpha\beta\gamma}_{kpq}&=&\frac{(k^2+p^2-q^2)c_p^{\alpha\beta}\delta_{\alpha\gamma}}
{[4k^2p^2-(k^2+p^2-q^2)^2]^{(3-d)/4}}\nonumber\\
&&+\frac{(k^2-p^2+q^2)c_q^{\alpha\gamma}\delta_{\alpha\beta}}{[4k^2p^2-(k^2+p^2-q^2)^2]^{(3-d)/4}}.
\end{eqnarray}
$c_k^{\alpha\beta}$ denote the direct correlation functions already introduced above
and $\Omega_d=2{\pi}^{d/2}/\Gamma(d/2)$ is the surface of a unit sphere in $d$ dimensions. $\Gamma(x)$ is
the gamma function.

The MCT Eqs.~(\ref{zwanzig_mori})-(\ref{vertex}) can be solved numerically, only. For this
we follow the standard procedure described in Ref.~\cite{Hajnal09} and discretize the wave numbers. This leads to a
finite, equally spaced grid of $K$ points, $k=(\hat{o}_d+\hat{k})\Delta k$ with $\hat{k}=0,1,\dots,K-1$ and $0<\hat{o}_d<1$.
The integrals in Eq.~(\ref{memory}) are then replaced by Riemann sums
\begin{equation}
\label{riemann}
\int_0^\infty \mathrm{d} p\int_{|k-p|}^{k+p} \mathrm{d} q \dots \mapsto (\Delta k)^2 \sum_{\hat{p}=0}^{K-1}  \sum_{\hat{q}=|\hat{k}-\hat{p}|}^{\min\{K-1,\hat{k}+\hat{p}\}}\dots
\end{equation}
and Eq.~(\ref{zwanzig_mori}) represents a
finite number of coupled equations.
For the offset we choose $\hat{o}_2=0.303$ \cite{Bayer07,Hajnal09}. 
The natural unit length is given by $1/\sqrt{n}$.
The choice $K=250$ and $\Delta k=0.2$ turns out to be sufficiently accurate to avoid larger discretization effects.

\subsection{Glass transition lines}

To locate the glass transition point, we introduce the nonergodicity parameters (NEPs) given by
$\boldsymbol{F}=\lim_{t\rightarrow\infty}\boldsymbol{\Phi}(t)$.
For our discretized model it can be proved \cite{Franosch02} that $\boldsymbol{F}$ is (with respect to $\succeq$)
the maximum real symmetric fixed point of
\begin{equation}
\label{fixed_point}
\boldsymbol{\mathcal{I}}[\boldsymbol{X}]=\boldsymbol{S}-(\boldsymbol{S}^{-1}+\boldsymbol{\mathcal{F}}[\boldsymbol{X},\boldsymbol{X}])^{-1}.
\end{equation}
$\boldsymbol{F}$ can be calculated numerically by iterating Eq.~(\ref{fixed_point}) starting with $\boldsymbol{X}=\boldsymbol{S}$.

For the binary dipole model with three independent control parameters, the glass transition takes place at a critical surface
where $\boldsymbol{F}$ jumps from $\boldsymbol{0}$ to $\boldsymbol{F}^c\succ\boldsymbol{0}$.
This surface can be represented as $\Gamma^c(x_s,\delta)$. Fixing $\delta=\delta^c$ defines a glass transition line (GTL).
A general expression for its slope has been derived in Ref.~\cite{Hajnal09}. Let $(\Gamma^c,x_s^c,\delta^c)$
be a critical point and $(\Delta\Gamma,\Delta x_s,\Delta\delta)=(\Gamma-\Gamma^c,x_s-x_s^c,\delta-\delta^c)$.
Then the following relation holds \cite{Hajnal09}
\begin{equation}
\label{slope}
\left.\frac{\partial \Gamma^c}{ \partial x_s}\right|_{(x_s,\delta)=(x_s^c,\delta^c)}
=-\left.\frac{\partial\sigma/\partial (\Delta x_s)}{\partial\sigma/\partial (\Delta\Gamma)}\right|
_{(\Delta\Gamma,\Delta x_s,\Delta\delta)=\vec{0}}
\end{equation}
where $\sigma$ is the so-called separation parameter \cite{Goetze09}.
Eq.~(\ref{slope}) can be evaluated in the so-called weak mixing limits $x_s\rightarrow 0$ and $x_b\rightarrow 0$ by making use of the
same perturbational approach as was performed for binary mixtures
of hard disks and hard spheres \cite{Hajnal09}. The few technical modifications which have to be taken into account
for the binary dipole model are summarized in \ref{wml}.
Then Eq.~(\ref{slope}) serves as a powerful tool for a fast and precise prediction of some
qualitative properties of the GTLs.

Let us also discuss some general properties of the GTLs. Our choice of $\Gamma$ (Eq.~(\ref{Gamma})) implies
\begin{equation}
\label{limits}
\Gamma^c(x_s,1)\equiv\Gamma^c(0,\delta)\equiv\Gamma^c(1,\delta)\equiv\Gamma_0^c
\end{equation}
where $\Gamma_0^c$ is the critical interaction parameter for the corresponding monodisperse system.
Further, an interchange of the roles of the big and small dipoles leads to the same physical scenario.
Therefore it is
\begin{equation}
\label{symmetry}
\Gamma^c(x_s,\delta)=\Gamma^c(1-x_s,1/\delta).
\end{equation}
Now, let us assume a small disparity in the susceptibilities, i.e. $0<(1-\delta)\ll 1$.
Then, with Eq.~(\ref{limits}) we can write
\begin{eqnarray}
\Gamma^c(x_s,\delta)&=&\Gamma_0^c+(1-\delta)\Gamma_1^c(x_s)+(1-\delta)^2\Gamma_2^c(x_s)\nonumber\\
& & +\mathcal{O}[(1-\delta)^3],
\label{symmetry3}
\end{eqnarray}
\begin{equation}
\label{koeff3}
\Gamma_i^c(0)=\Gamma_i^c(1)=0,\quad i=1,2.
\end{equation}
Eq.~(\ref{symmetry}) implies for the coefficient functions
\begin{equation}
\label{koeff1}
\Gamma_1^c(x_s)=-\Gamma_1^c(1-x_s),
\end{equation}
\begin{equation}
\label{koeff2}
\Gamma_2^c(x_s)=\Gamma_2^c(1-x_s)-\Gamma_1^c(1-x_s).
\end{equation}
We can conclude: if $\Gamma_1^c(x_s)$ would be a non-vanishing function, then to leading order in $(1-\delta)$
we would obtain 
GTLs which, after subtracting $\Gamma_0^c$, are antisymmetric with respect to the equimolar composition $x_s=1/2$. We could write
$(\Gamma^c(x_s,\delta)-\Gamma_0^c)\cong-(\Gamma^c(1-x_s,\delta)-\Gamma_0^c)$ and $\Gamma^c(1/2,\delta)\cong\Gamma_0^c$.
If, however, there holds
\begin{equation}
\label{zero}
\Gamma_1^c(x_s)=0 
\end{equation}
for all $x_s$ while $\Gamma_2^c(x_s)$ is a non-vanishing function, then Eqs.~(\ref{symmetry3}) and (\ref{koeff2}) imply
\begin{equation}
\label{symmetry4}
\Gamma^c(x_s,\delta)\cong\Gamma^c(1-x_s,\delta)
\end{equation}
which means that the GTLs become symmetric with respect to $x_s=1/2$ in the limit of small disparity in the
susceptibilities. There holds also the (somewhat weaker) inverse statement. If Eq.~(\ref{symmetry4}) is assumed to be true for $(1-\delta)\ll 1$, then
Eq.~(\ref{koeff1}) implies Eq.~(\ref{zero}). For the slope $s^c(x_s,\delta)=(\partial\Gamma^c/\partial x_s)(x_s,\delta)$ of a GTL we can formulate
further special (numerically testable) conclusions from Eq.~(\ref{zero}):
\begin{equation}
\label{symmetry5}
(\partial s^c/\partial \delta)(0,1)=(\partial s^c/\partial \delta)(1,1)=0.
\end{equation}
Since the $T/2$-HNC closure relation has no free adjustable parameters,
their structure factors will be taken for the calculation of the slope (\ref{slope}). We will be able to produce numerical results with sufficiently high
accuracy to demonstrate that Eqs.~(\ref{symmetry4}) and (\ref{symmetry5}), and thus also Eq.~(\ref{zero}), are indeed valid for the
binary dipole model.

\section{MCT results and discussion}

\label{results}

\subsection{Glass transition lines}

\begin{figure}
\includegraphics[width=1\columnwidth]{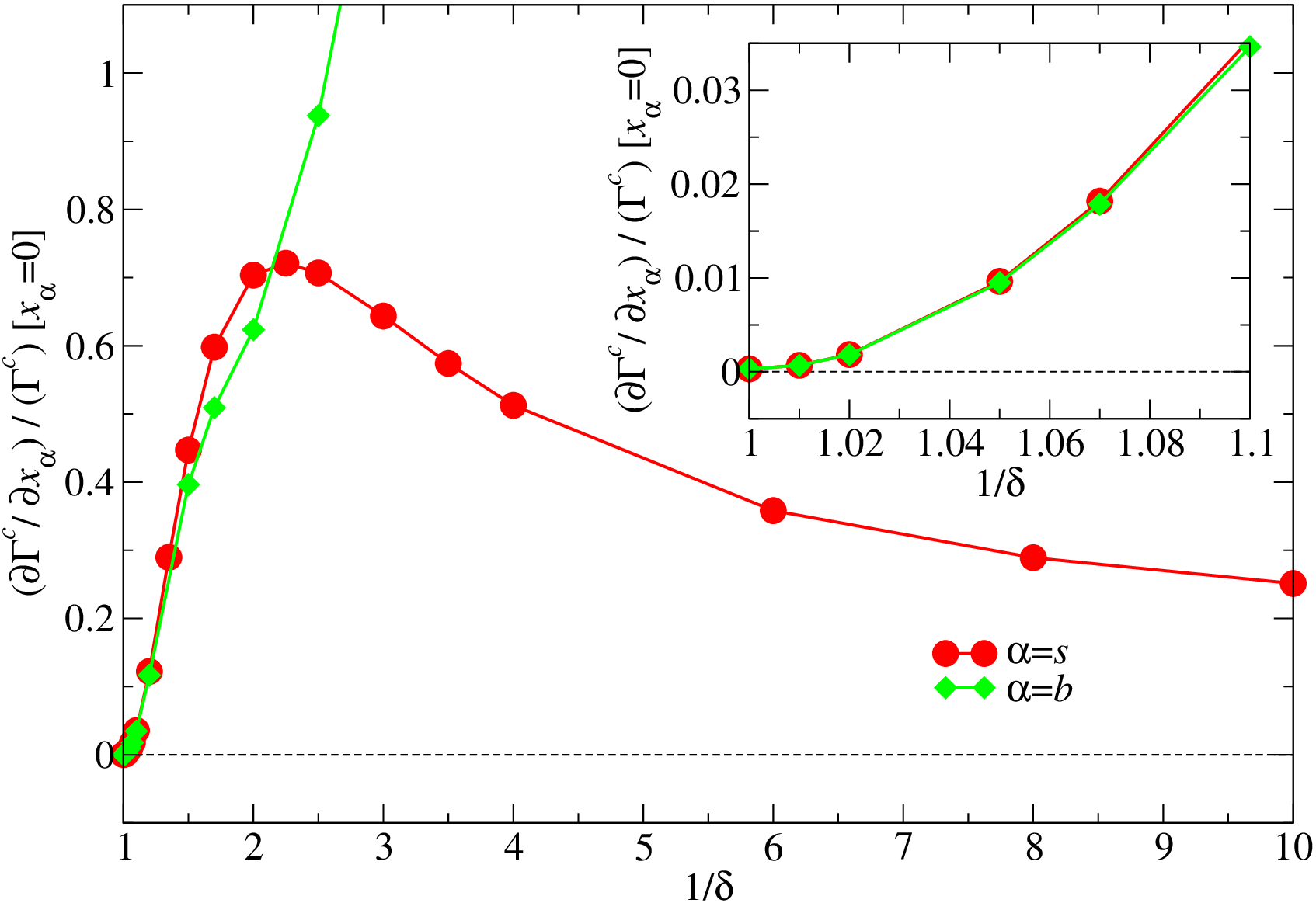}
\caption{(Color online) Normalized slopes of the GTLs
for binary mixtures of dipolar particles in 2D at $x_s=0$ and $x_b=0$
calculated by using Eq.~(\ref{slope}) and the $T/2$-HNC closure given by Eq.~(\ref{bridge2}).
It is $\Gamma_0^c\cong 97.704$ for the monodisperse limit. The inset presents the slopes on a much
finer scale very close to $\delta=1$.}
\label{fig_slope}
\end{figure}

Fig.~\ref{fig_slope} shows normalized slopes $(\partial\Gamma^c/\partial x_{\alpha})(\Gamma^c)^{-1}$
of the GTLs at $x_\alpha=0$, $\alpha=s,b$, as functions
of $1/\delta$ calculated by using Eq.~(\ref{slope}) and the $T/2$-HNC closure given by Eq.~(\ref{bridge2}).
Because the case $\delta=1$ is equivalent to a one-component model,
the slopes have to be zero at this point. Our numerical results clearly support this statement (see inset of Fig.~\ref{fig_slope}).
We observe that for $0<\delta<1$ the slopes become always positive.
This means that, compared to the corresponding one-component liquid, the presence of a small concentration of a second kind of
particles \textit{always} increases the critical interaction parameter $\Gamma^c(x_s,\delta)$ at which vitrification sets in.
For fixed temperature $T$ and fixed total particle density $n$ this implies that,
compared to a given one-component system, the presence of a small concentration of a second kind of dipolar particles always increases
the critical field $B^c$ and accordingly
the critical value $m^c$ of the average magnetization per particle at which the system vitrifies.
In this sense, mixing stabilizes the liquid state which is also called plasticization,
{especially in engineering literature, see Ref.~\cite{Chen} for example.}

Now, let us focus on the $\delta$-dependence of the slope curve for $x_s=0$. Starting from zero at $\delta=1$, the slope first increases upon decreasing $\delta$.
After exhibiting a maximum at $\delta_+\cong0.444$, the slope monotonically decays for asymptotically small $\delta$.
For $x_b=0$ the slope again starts from zero at $\delta=1$ and increases strongly upon decreasing $\delta$.
As a further result,
the inset in Fig.~\ref{fig_slope} demonstrates that the slope curves for $x_s=0$ and $x_b=0$ become
identical for $(1-\delta)\ll1$. This is already a first hint on the validity of Eq.~(\ref{symmetry4}),
i.e. on the symmetry of the GTLs with respect to $x_s=1/2$ for $(1-\delta)\ll1$.
Furthermore, the numerical data indicate that the slopes, i.e. the derivatives
of both slope curves vanish at $\delta=1$, i.e. the data indicate that Eq.~(\ref{symmetry5}) holds. 

The results in Fig.~\ref{fig_slope} already allow us to predict some properties of the GTLs. Both, $x_s=0$ and $x_s=1$,
define one-component models with the same critical interaction parameter $\Gamma_0^c$. Hence, all GTLs must exhibit at least one
maximum, since $(\partial\Gamma^c/\partial x_s)(x_s=0,\delta)>0$ and $(\partial\Gamma^c/\partial x_s)(x_s=1,\delta)
=-(\partial\Gamma^c/\partial x_b)(x_b=0,\delta)<0$ for $0<\delta<1$.
Next, if for two values, $\delta_1$ and $\delta_2$, it is
$(\partial\Gamma^c/\partial x_s)(x_s=0,\delta_1)<(\partial\Gamma^c/\partial x_s)(x_s=0,\delta_2)$
\textit{and} 
$(\partial\Gamma^c/\partial x_b)(x_b=0,\delta_1)>(\partial\Gamma^c/\partial x_b)(x_b=0,\delta_2)$,
i.e.
$(\partial\Gamma^c/\partial x_s)(x_s=1,\delta_1)<(\partial\Gamma^c/\partial x_s)(x_s=1,\delta_2)$,
then the GTLs for $\delta_1$ and $\delta_2$ have an odd number of intersection points and therefore at least one crossing.
Otherwise the number of intersection points is even.
Since the slope curve $s^c(x_s,\delta)=(\partial\Gamma^c/\partial x_s)(x_s,\delta)$
at $x_s=0$ decreases monotonically with decreasing $\delta$ for $0<\delta<\delta_+$
while $-s^c(x_s,\delta)$
increases monotonically with decreasing $\delta$
for all $0<\delta<1$ at $x_s=1$, two GTLs corresponding to $\delta_1$ and $\delta_2$ must have an odd number of crossings, if
$0<\delta_1<\delta_2<\delta_+$.

In the following, we present GTLs calculated both by using $T/2$-HNC and MC structure factors as input for MCT.
Due to crystallization effects, the control parameters for the MC data had to be restricted to $0.1\leq x_s\leq0.9$
and $0<\delta\leq0.6$.

\begin{figure}
\includegraphics[width=1\columnwidth]{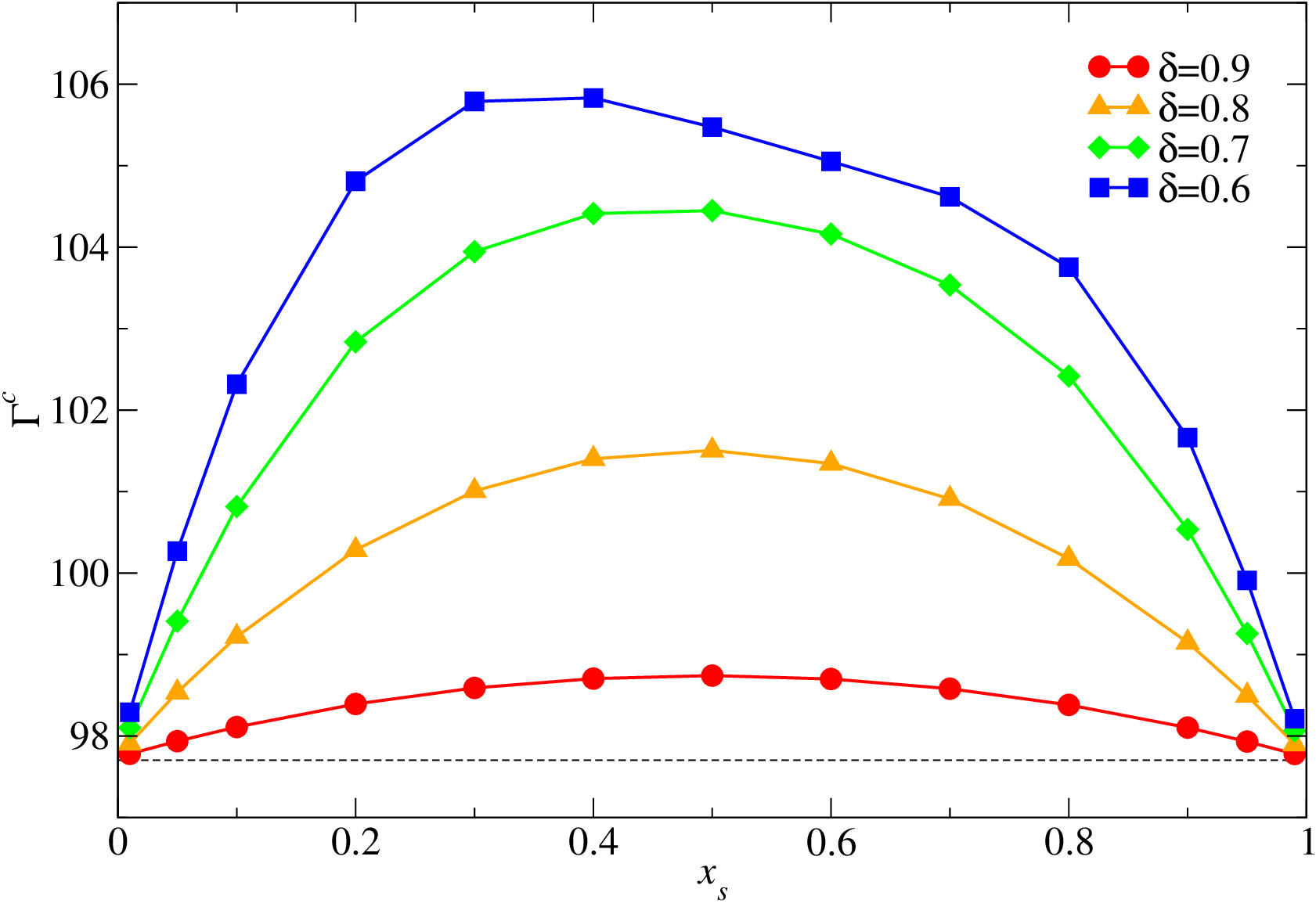}
\caption{(Color online) Glass transition lines for binary mixtures of dipolar particles in 2D
with moderate susceptibility ratios 
calculated by using the $T/2$-HNC closure.
$\Gamma_0^c$ is indicated by the dashed line.}
\label{fig_phase1}
\end{figure}
 
Fig.~\ref{fig_phase1} shows GTLs for moderate susceptibility ratios $\delta_+<0.6\leq\delta\leq0.9$ calculated by using $T/2$-HNC
structure factors as input for MCT.
For $0<x_s<1$, all GTLs are strictly above the monodisperse value $\Gamma_0^c$ and exhibit a \textit{single} maximum. In addition,
the GTLs do not intersect. The three lines for $0.7\leq\delta\leq0.9$ are almost symmetric with respect to the equimolar
composition $x_s=1/2$, i.e. the data strongly support Eq.~(\ref{symmetry4}) and thus also Eq.~(\ref{zero}).
All these properties are compatible with the results for the slopes shown in Fig.~\ref{fig_slope}.

\begin{figure}
\includegraphics[width=1\columnwidth]{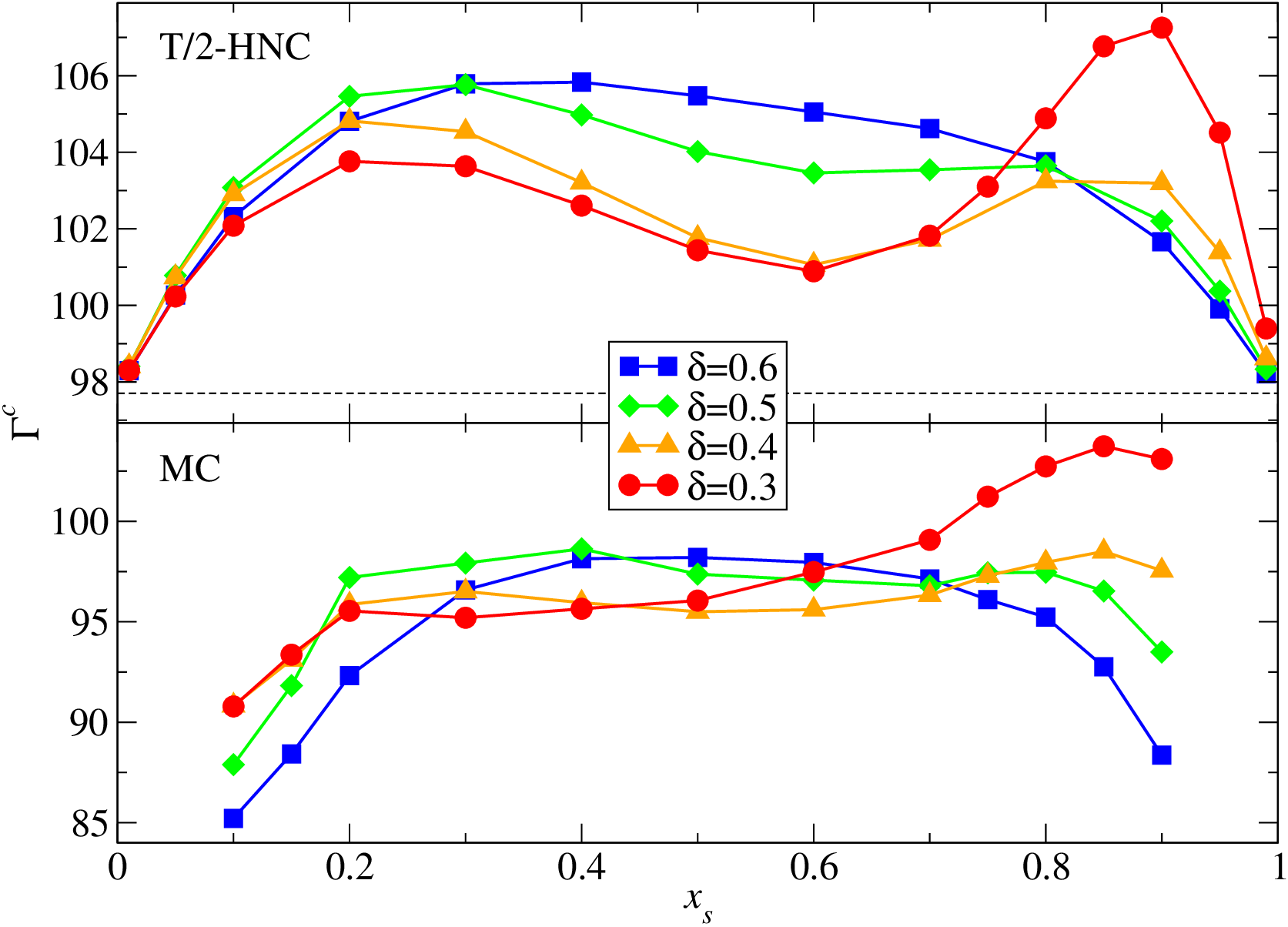}
\caption{(Color online) Glass transition lines for binary mixtures of dipolar particles in 2D
calculated by using $T/2$-HNC structure factors (upper panel)
and MC structure factors (lower panel).
The dashed line in the upper panel indicates the $T/2$-HNC value for
$\Gamma_0^c$.}
\label{fig_phase2}
\end{figure}

Fig.~\ref{fig_phase2} presents GTLs for $0.3\leq\delta\leq0.6$ calculated both by using $T/2$-HNC and MC structure factors.
The $T/2$-HNC curves again are strictly above the monodisperse value $\Gamma_0^c$ for $0<x_s<1$.
For the MC curves, we can not test this statement strictly since we have no reasonable estimate for $\Gamma_0^c$ due to
crystallization effects. Therefore, we focus first on the $T/2$-HNC results. While the GTL for $\delta=0.6$ exhibits only
a single rather flat maximum, the other GTLs with $0.3\leq\delta\leq0.5$ develop
two local maxima with a local minimum at $x_s\approx0.6$ in between.
Due to the occurrence of this local minimum for $\delta=0.5$
\textit{and} the increase of both slope curves $s^c(x_s,\delta)$ at $x_s=0$ and $-s^c(x_s,\delta)$
at $x_s=1$ with decreasing $\delta$ for $\delta_+<\delta<1$, the curves for $\delta=0.5$ and $\delta=0.6$
have two intersection points within the interval $0<x_s<1$.
If, however, we plot a pair of the shown GTLs with $0.3\leq\delta_1<\delta_2\leq0.5$, then
we obtain exactly one intersection point within the interval $0<x_s<1$.
For the pair $\delta_1=0.3$ and $\delta_2=0.4$
the existence of an odd number of intersection points is enforced by the facts
$|(\partial\Gamma^c/\partial x_s)(x_s=0,\delta_1)|<|(\partial\Gamma^c/\partial x_s)(x_s=0,\delta_2)|$ and
$|(\partial\Gamma^c/\partial x_s)(x_s=1,\delta_1)|<|(\partial\Gamma^c/\partial x_s)(x_s=1,\delta_2)|$
following both from the GTLs by numerical differentiation and 
also from the slopes in Fig.~\ref{fig_slope}, since $\delta_i<\delta_+$ for $i=1,2$.
The upper panel of Fig.~\ref{fig_phase2} shows that only one intersection point occurs for the two GTLs.
Now, let us consider the MC results in Fig.~\ref{fig_phase2}.
The data are numerically less precise and the occurring extrema in the curves are not as clearly pronounced as
for the $T/2$-HNC results. Furthermore, the MC curves do not exhibit the same number of intersection points
as the corresponding $T/2$-HNC ones.
Nevertheless, the MC data clearly support at least the qualitative $x_s$-dependencies of
the GTLs predicted by the $T/2$-HNC closure.
Note that the scales for $\Gamma^c$ and its total variation are also similar in both approaches.

\begin{figure}
\includegraphics[width=1\columnwidth]{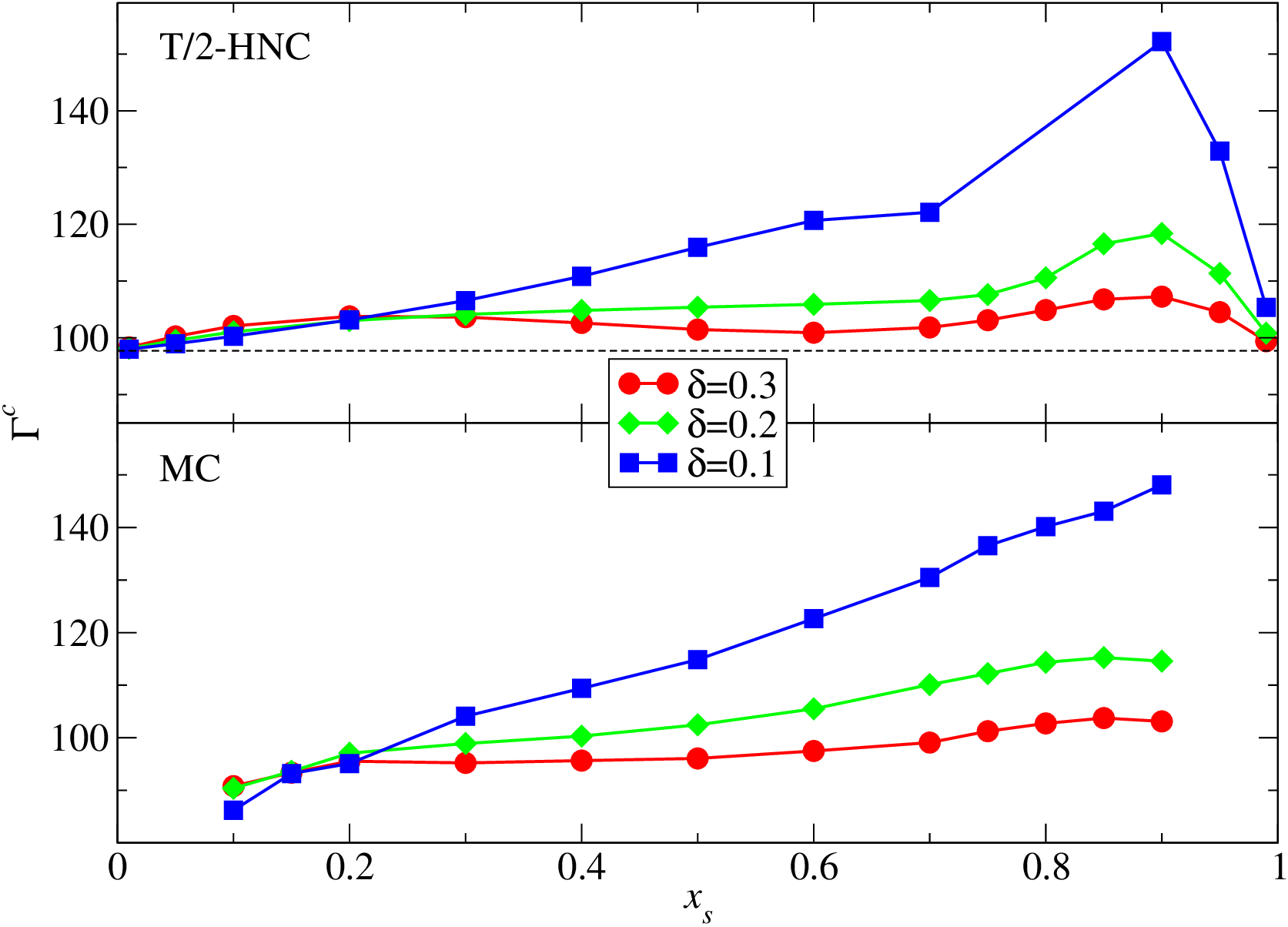}
\caption{(Color online) Glass transition lines for binary mixtures of dipolar particles in 2D
with larger disparities in the susceptibilities
calculated by using $T/2$-HNC structure factors (upper panel)
and MC structure factors (lower panel).
The dashed line in the upper panel indicates the $T/2$-HNC value for
$\Gamma_0^c$. In the neighborhood of $x_s=0.8$ and $\delta=0.1$ we had problems in finding stable numerical solutions for
the $T/2$-HNC structure factors, and thus there are no data points for $\Gamma^c$ in this region in the upper panel.}
\label{fig_phase3}
\end{figure}

Fig.~\ref{fig_phase3} shows GTLs
for larger disparities in the susceptibilities, $0.1\leq\delta\leq0.3$,
calculated both by using $T/2$-HNC and MC structure factors as input for MCT.
Let us first focus on the $T/2$-HNC data.
Again, all GTLs are strictly above the monodisperse value $\Gamma_0^c$ for $0<x_s<1$.
The GTL for $\delta=0.3$, which is also shown in Fig.~\ref{fig_phase2} on a finer scale for $\Gamma^c$,
exhibits two local maxima at $x_s\approx0.25$ and $x_s\approx0.88$ and in between a clearly
pronounced local minimum at $x_s\approx0.6$
(we have obtained these numbers via spline-interpolations).
Upon decreasing $\delta$, the amplitude of the maximum at $x_s\approx0.88$ strongly increases while the maximum
at $x_s\approx0.25$ and thus also the minimum at $x_s\approx0.6$ become less pronounced.
As a result, the GTL for $\delta=0.2$ exhibits only a single maximum at $\delta\approx0.89$. The GTL for $\delta=0.1$ seems to be
qualitatively similar, but 
in the neighborhood of $x_s=0.8$ and $\delta=0.1$ we had problems in finding stable numerical solutions for
the $T/2$-HNC structure factors, and thus we have no data points for $\Gamma^c$ in this region.
As already predicted by the slopes presented in Fig.~\ref{fig_slope}, each pair of the shown GTLs exhibits an intersection point within
the interval $0<x_s<1$.
The MC results in Fig.~\ref{fig_phase3} clearly support the trends predicted by the $T/2$-HNC closure.
Note especially the similarity of the intersection points of the curves.
Further, both approaches yield almost identical scales for $\Gamma^c$ and its total variation.

The common feature of all $T/2$-HNC results in Figs.~\ref{fig_slope} - \ref{fig_phase3} is the fact that there holds strictly
\begin{equation}
\label{plastic}
\Gamma^c(x_s,\delta)>\Gamma_0^c
\end{equation}
for all tested $0<\delta<1$ and $0<x_s<1$.
The corresponding MC results in Figs.~\ref{fig_phase2} and \ref{fig_phase3} are also consistent with these findings.
Thus we can conclude: for the binary dipole model in 2D, MCT predicts always a stabilization of the liquid state due to mixing. Further, the amplitude of this
plasticization effect is quite large. For instance, the variation of $\Gamma^c$ in Fig.~\ref{fig_phase2} is of the order of $10$ percent.
{The GTLs for $\delta=0.1$ (Fig.~\ref{fig_phase3}) vary by even more than $50$ percent.
Tests of MCT in the last two decades have shown that for model systems with smooth pair potentials 
the critical temperatures predicted by MCT differ from those determined from computer simulations
by about a factor of two, i.e. $100$ percent \cite{Goetze09}.
This does not necessarily imply for the binary dipole model that the $50$ percent variation of $\Gamma^c$ for $\delta=0.1$
can not be observed experimentally. If $\Gamma^c_{exp}(x_s,\delta=0.1)\approx2\Gamma^c(x_s,\delta=0.1)$ for $0<x_s<1$,
then the experimentally determined critical vale $\Gamma^c_{exp}$ would vary by about $50$ percent, as well.
Thus, the predicted plasticization effect should be experimentally testable.}

\subsection{Comparison to experiments}

\begin{figure}
\includegraphics[width=1\columnwidth]{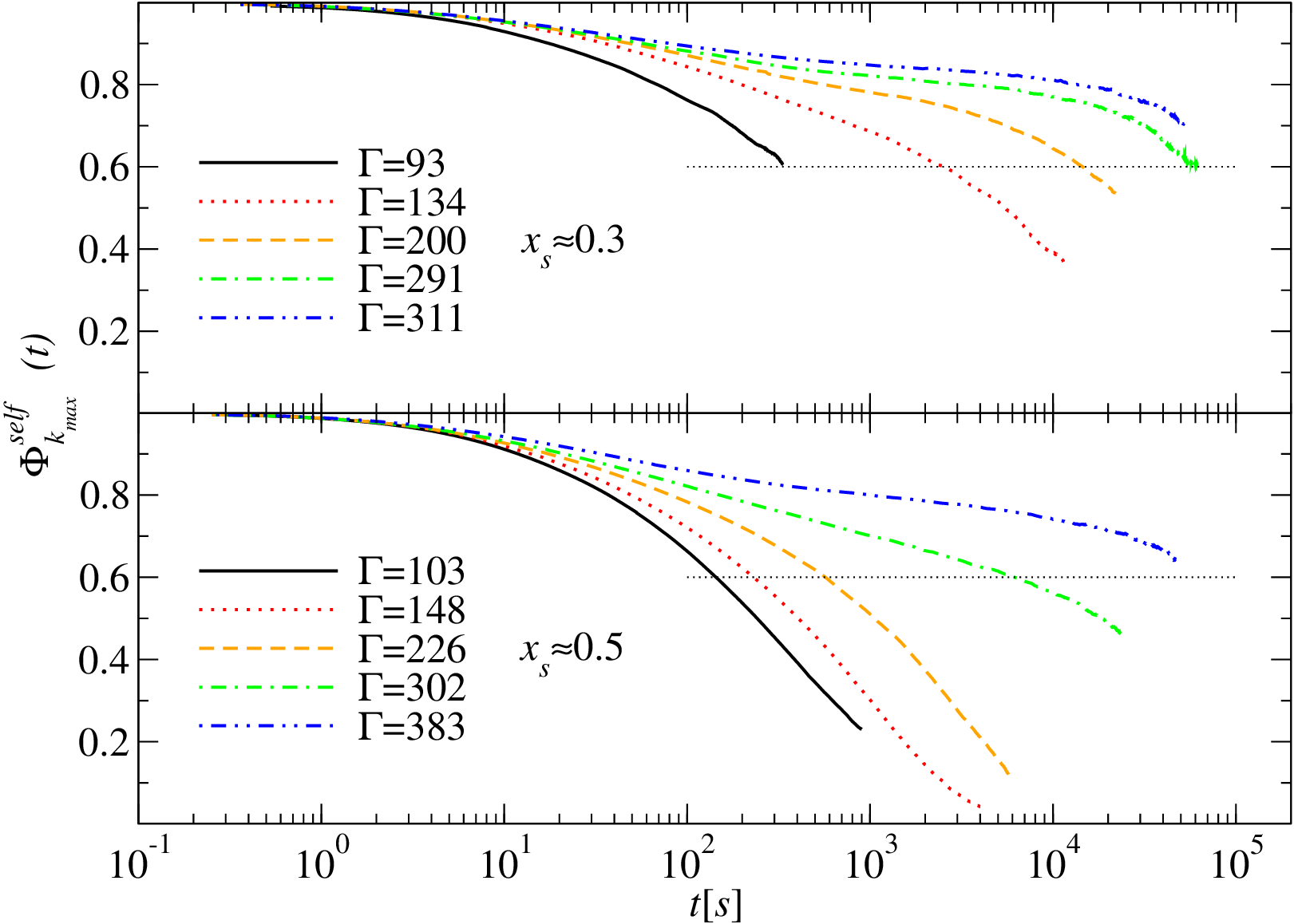}
\caption{(Color online) Self-intermediate scattering functions of K\"onig et al. \cite{Koenig05}
measured experimentally at $\delta\approx0.1$ and $x_s\approx0.3$ (upper panel) and $x_s\approx0.5$ (lower panel).
The functions were averaged over big and small particles.
$k_{max}$ denotes the wave number corresponding to the first maximum of $g^{bb}(r)$. See Ref.~\cite{Koenig05}
for details. The dotted horizontal lines mark the value $0.6$ (see text).}
\label{scatter}
\end{figure}

Of course, it would be interesting to determine $\Gamma^c(x_s,\delta)$ experimentally, which has not been done so far.
Another possibility to check our prediction, as already discussed for binary mixtures of hard disks and hard spheres
in Refs.~\cite{Hajnal09,Voigtmann03}, is the strong variation in the $\alpha$-relaxation
time scale $(\boldsymbol{\tau}^{rel})_k^{\alpha\beta}$ upon changing $x_s$ or $\delta$.
Let us fix $\delta=0.1$ (which is approximately the experimental value \cite{Koenig05}) and choose two compositions $x_s^{(1)}=0.3$ and $x_s^{(2)}=0.5$. Now, let us further fix $\Gamma$
such that $0<[\Gamma^c(x_s^{(i)},\delta)-\Gamma]/\Gamma\ll1$ for $i=1,2$.
With the asymptotic scaling law $(\boldsymbol{\tau}^{rel})^{\alpha\beta}_k\sim([\Gamma^c(x_s,\delta)-\Gamma]/\Gamma)^{-\gamma(x_s,\delta)}$
\cite{Goetze09} and the fact $\Gamma^c(x_s^{(1)},\delta)<\Gamma^c(x_s^{(2)},\delta)$ (see Fig.~\ref{fig_phase3})
we obtain $(\boldsymbol{\tau}^{rel})_k^{\alpha\beta}(\Gamma,x_s^{(2)},\delta)\ll(\boldsymbol{\tau}^{rel})_k^{\alpha\beta}(\Gamma,x_s^{(1)},\delta)$.
Thus, we expect for this chosen value of $\Gamma$ and $\delta=0.1$ that binary mixtures with $x_s=x_s^{(2)}=0.5$ will behave more liquid-like than
corresponding mixtures with $x_s=x_s^{(1)}=0.3$, i.e. their structural relaxation is faster.

{Next we consider the experimental results of K\"onig et al. \cite{Koenig05}
for the self-intermediate density correlators shown in Fig.~\ref{scatter}
which were obtained for $\delta\approx0.1$ and two different compositions $x_s\approx0.3$ and $x_s\approx0.5$.
Note that the experimental values for $x_s$ of K\"onig et al. \cite{Koenig05} are not as accurate as the corresponding ones
of Ebert \cite{ebert_phd} (see caption of our Fig.~\ref{fig_mc}).
From Fig.~\ref{fig_phase3} we find $\Gamma^c(x_s=0.3,\delta=0.1)=104<\Gamma^c(x_s=0.5,\delta=0.1)=115$ in case of the static
input from our MC simulation. On the other hand, for the structural relaxation time $\tau(\Gamma,x_s,\delta)$
defined by $\Phi^{self}_{k_{max}}(\tau,\Gamma,x_s,\delta)=0.6$ we can read off from Fig.~\ref{scatter} that
$\tau_{exp}(\Gamma=226,x_s\approx0.5,\delta\approx0.1)=569s\ll\tau_{exp}(\Gamma=200,x_s\approx0.3,\delta\approx0.1)=14513s$.
Since $\tau(\Gamma_1,x_s,\delta)<\tau(\Gamma_2,x_s,\delta)$ for $\Gamma_1<\Gamma_2<\Gamma^c(x_s,\delta)$, it must be
\begin{eqnarray}
&\tau_{exp}(\Gamma=200,x_s\approx0.5,\delta\approx0.1)\ll\tau_{exp}(\Gamma=\nonumber\\
&200,x_s\approx0.3,\delta\approx0.1).
\label{ineq}
\end{eqnarray}
Let $\Gamma^c_{exp}(x_s,\delta)$ be the critical value at which an extrapolation of $\tau_{exp}(\Gamma,x_s,\delta)$
diverges as a power law $\tau_{exp}(\Gamma,x_s,\delta)\sim[\Gamma^c_{exp}(x_s,\delta)-\Gamma]^{-\gamma(x_s,\delta)}$.
Assuming that the exponent and the proportionality factor of that
power law do not change sensitively under a change of $x_s\approx0.3$ to $x_s\approx0.5$ (for fixed $\delta\approx0.1$),
then the inequality (\ref{ineq}) implies that the distance of $\Gamma=200$ to $\Gamma^c_{exp}(x_s\approx0.3,\delta\approx0.1)$
is smaller than the distance to $\Gamma^c_{exp}(x_s\approx0.5,\delta\approx0.1)$.
Consequently, we have $\Gamma^c_{exp}(x_s\approx0.3,\delta\approx0.1)<\Gamma^c_{exp}(x_s\approx0.5,\delta\approx0.1)$, in agreement
with the MCT predictions above. This inequality between the experimental values for $\Gamma^c_{exp}$ is also consistent with the fact that
the mixture with $x_s\approx0.3$ and $\Gamma=200$ already shows a plateau whereas the mixture with $x_s\approx0.5$ and $\Gamma=226$ does not.
}

This first attempt to compare our results with the experimental ones
indicates that MCT is able to predict at least qualitatively the change in the relaxation
behavior upon composition change from $x_s\approx 0.3$ to $x_s\approx 0.5$.
Of course, it is necessary to strongly extend the parameter range for $x_s$, but also for $\delta$, to test the validity
of our MCT predictions more systematically.
Performing such a systematic test has also to account for the
overestimation of the temperature scale at which vitrification sets in.
For $\delta=0.1$ and $x_s=0.3$ MCT yields $\Gamma^c\approx105$
by both using $T/2$-HNC and MC structure factors as input.
In the upper panel of Fig.~\ref{scatter}, however, the development of a plateau as a precursor of a possible glass
transition only becomes clearly visible in the experimental data for $\Gamma\geq200$.
Hence, MCT overestimates the scale for the glass transition temperature by about a factor of two,
as well-known for other systems.

\section{Summary and conclusions}

\label{conclusions}

The major motivation of our study has been to explore the effect of mixing on the ideal glass transition of a binary mixture
of dipolar colloids in 2D and the comparison with corresponding experimental data. The calculations were done in the
framework of MCT.

The MCT equations were solved by using the static structure factors from MC simulations and integral equation theory.
For the latter we have used the $T/2$-HNC closure relation because it does not involve adjustable parameters and
fits the MC results for the partial static structure factors rather well. The very good agreement of the partial static structure factors from experiment \cite{Koenig05}
with those from our MC simulation using dipolar point particles has demonstrated that the dipolar colloidal system is well modeled by point particles
in 2D interacting only via the induced dipole potentials (\ref{potential}), at least in the parameter range where the
glass transition occurs.

The main result we have found is that the effect of mixing on the glass transition depends strongly on the liquid system.
In contrast to binary mixtures of hard disks and hard spheres \cite{Hajnal09,Voigtmann03},
MCT for the binary liquid of point dipoles
predicts always a stabilization
of the liquid state due to mixing.
More explicitly, for fixed temperature $T$ and fixed total particle density $n$
the critical magnetic field $B^c$ and accordingly the critical value $m^c$ of the magnetization
per particle at which the binary dipolar liquid vitrifies is always larger, compared to a one-component system.
The amplitude of this predicted plasticization effect
is quite large, the GTLs for $\delta=0.1$ shown in Fig.~\ref{fig_phase3} vary by more than $50$ percent.

The $\alpha$-relaxation times of the experimentally obtained self-intermediate density correlators shown in
Fig.~\ref{scatter} for $x_s\approx0.3$ and $x_s\approx0.5$ clearly support the predicted plasticization effect.
This hints that MCT may be able to predict qualitatively the change in the relaxation
behavior upon composition changes. The experimental data also show
that MCT {systematically} overestimates the scale for the glass transition temperature by more than a factor of two,
which is a well-known phenomenon of MCT \cite{Goetze09}.

\begin{figure}
\includegraphics[width=1\columnwidth]{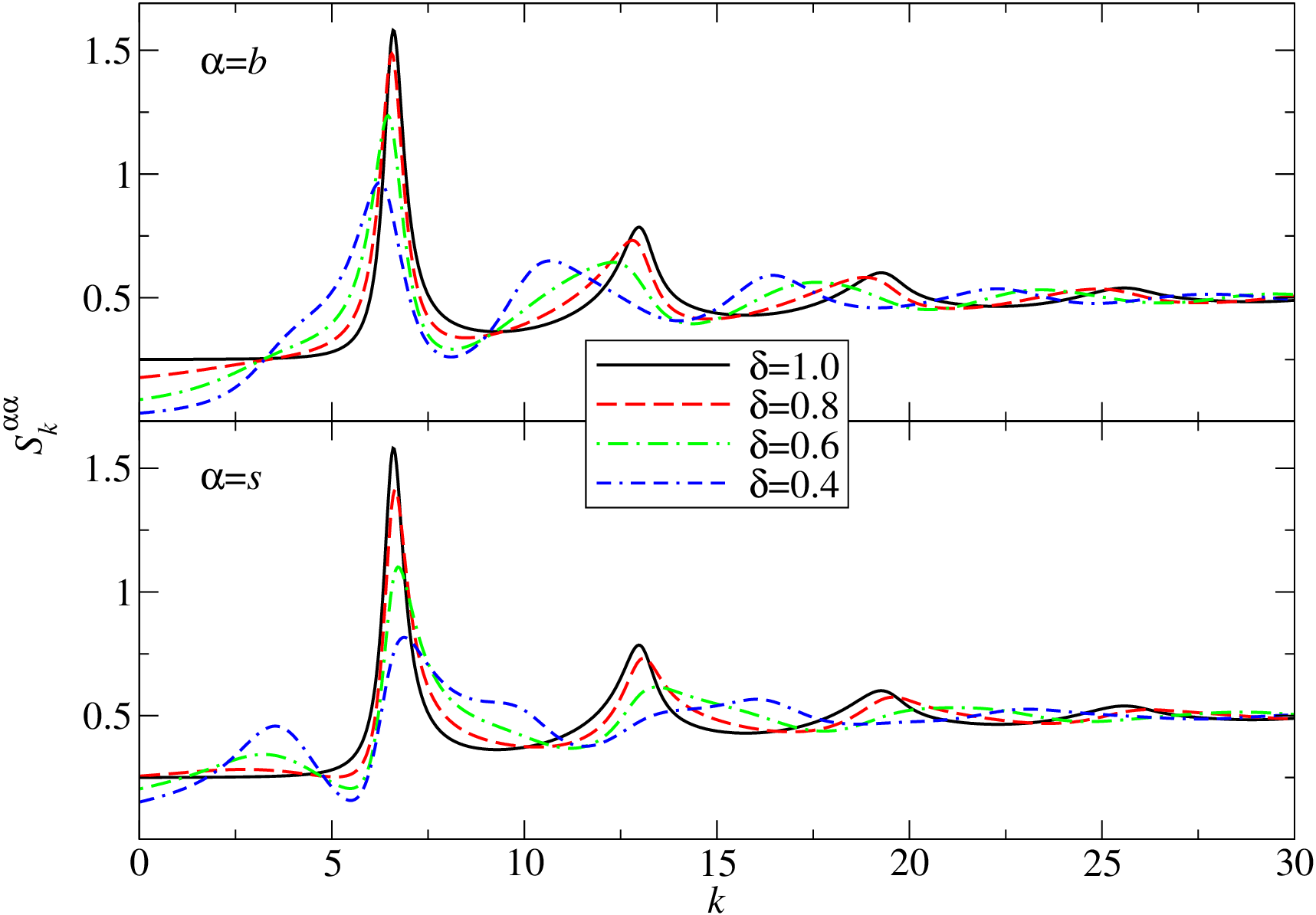}
\caption{(Color online)
Partial structure factors $S_k^{\alpha\alpha}$ for binary mixtures of dipolar particles in 2D
at $\Gamma=95$ and $x_s=0.5$ calculated by using the $T/2$-HNC closure.}
\label{fig_cluster}
\end{figure}

{At this point, let us also speculate about the physical origin of the predicted plasticization
effect. As pointed out by Hoffmann et al. \cite{Hoffmann06}, the smaller particles in the binary dipole model tend to form
clusters whereas the bigger particles remain more homogeneously distributed in space.
This partial clustering manifests itself by the occurrence of a prepeak in the partial structure factors $S_k^{ss}$
of the smaller particles at low $k$. Furthermore, the cited clustering effect seems to occur as a generic feature of the binary
dipole model: all partial structure factors $S_k^{ss}$ with $0<\delta<1$ and $0<x_s<1$ used for calculating the glass transition lines in
Figs.~\ref{fig_phase1}-\ref{fig_phase3} exhibit a prepeak. On the other hand, MCT predicts plasticization for all $0<\delta<1$ and $0<x_s<1$
where the only model specific input to MCT are the partial structure factors $S_k^{\alpha\beta}$.
Thus, it could be that the partial clustering of the smaller particles is the physical origin for the plasticization effect
predicted by MCT.}

{To investigate this conjecture in more detail, we have calculated $T/2$-HNC structure factors $S_k^{\alpha\beta}$ for $\Gamma=95$, $x_s=0.5$,
and $\delta=1.0$, $0.8$, $0.6$, and $0.4$ from which the $bb$ and $ss$ elements are shown in Fig.~\ref{fig_cluster}. The case $\delta=1.0$
is a monodisperse system close to its MCT glass transition point which shows no clustering and by definition satisfies
$S_k^{bb}=S_k^{ss}$. Now, by decreasing $\delta$ we observe five qualitative effects on $S_k^{\alpha\alpha}$:
(i) a growing prepeak in $S_k^{ss}$ at $2<k<5$, (ii) a decrease in $S_k^{bb}$ for $0<k<3$,
(iii) a decrease in the amplitude in the main- and the subsequent peaks in both $S_k^{bb}$ and $S_k^{ss}$,
(iv) a shrinking of the $k$ scale for the oscillations of $S_k^{bb}$, and (v) a stretching
of the $k$ scale for the oscillations of $S_k^{ss}$.
The plasticization effect predicted by MCT must be a result from an interplay of these five effects.
Since these effects are competing, no definite conclusion can be drawn concerning the plasticization effect.
Thus, at least from our analysis here
it is not obvious whether there is a connection between clustering and plasticization or not.
}

We have also discussed some general properties of the glass transition diagram and its GTLs. For small
disparities in the susceptibilities, i.e. for $\delta$ close to one, the GTLs are symmetric like for
binary hard disks \cite{Hajnal09,Voigtmann03}. Two GTLs with $\delta_1$ and $\delta_2$ below $\delta_+\approx0.444$
must have an odd number of intersection points. This prediction is only based on the application of the
analytical expression for the slope $(\partial\Gamma^c/\partial x_s)(x_s,\delta)$ at $x_s=0$ and $x_s=1$.
Furthermore, depending on $\delta$ the GTLs exhibit either one maximum, for $\delta$ closer to one or 
$\delta$ very small, and two maxima with a minimum in between, for intermediate values of $\delta$.
This behavior also differs qualitatively from that of binary hard disks in 2D and binary hard spheres in 3D.

As a side result to be mentioned,
with the $T/2$-HNC closure we have suggested a new bridge function approximation
which is useful for calculating the static structure functions of dipolar systems. Of course,
our finding is only of empirical nature.

The various properties of the GTLs of binary liquids of point dipoles in 2D predicted here in the framework of MCT
are a challenge to be tested in more detailed numerical simulations and by extending the present experimental
approach. But the theoretical approach can be extended as well. For instance, for the present model it will be
interesting to study
the fully time-dependent density correlators.
A further interesting topic will be the discussion of the glass
transition behavior of binary hard disks with dipolar interactions at intermediate and high packing fractions,
where the mixing effects for dipolar point particles  and those for hard disks \cite{Hajnal09} may interfere.
Such investigations are work in progress and will be published elsewhere.

\section*{Acknowledgments}

We thank J.~M.~Brader, F. Ebert, M.~Fuchs, P.~Keim, S.~H.~L.~Klapp, G.~Maret and Th.~Voigtmann
for stimulating and helpful discussions. We especially thank F.~Ebert and H.~K\"onig for providing their
experimental data for the static structure factors and for the self-intermediate scattering functions, respectively.
The latter which became possible due to the support by G.~Maret has
enabled us for a first
test of our MCT predictions.

\appendix

\section{Monte Carlo simulation}
\label{mc}

We simulate $N=1600$ particles within a cubic box of dimensionless length $L=\sqrt N=40$
with the standard periodic boundary conditions for particle motion.
This corresponds to setting the total particle number density to 1 since
the magnetic field $B$ and the susceptibility ratio $\delta$ are the only physical control
parameters, see Eq.~(\ref{Gamma}). 
We arrange $8$ virtual copies of the simulation box around it.
These so-called image boxes contain at every simulation step exactly the same
particle configurations as the main simulation box.
To calculate the potential energy of a particle inside the main simulation box,
the interactions with all particles within the main and the $8$ image boxes are
taken into account. This is necessary due to the long-ranged $1/r^3$ pair potentials.
However, these potentials are absolutely integrable in 2D and thus
we can calculate the potential energies without advanced techniques like Ewald summation.
A trial move for a chosen particle consists of a local random displacement
$\Delta x, \Delta y \in[-\varepsilon,\varepsilon]$, $\varepsilon=0.15$ of the considered particle (and also of its copies
in the image boxes) followed by calculating the potential energy cost $\Delta u$
for this displacement and choosing a random number $z\in[0,1]$. The trial move is
accepted if $z\leq\exp(-\Delta u/k_BT)$, and rejected, otherwise. Typical acceptance probabilities are
between $0.3$ and $0.5$.
A so-called sweep
is a sequence consisting of exactly one trial move for every particle in the main simulation
box.

The calculation of the static correlation matrix $\boldsymbol{S}_k$ consists of four steps.
First, we randomly distribute the $N=1600$ particles.
Second, we perform $10000$ sweeps for equilibrating the system.
As third step, we perform $20000$ sweeps to record a histogram for
the radial distribution functions
$g^{\alpha\beta}(r)$. For this purpose, $r$ is discretized
as a Lado grid \cite{Lado}
with a real space cutoff $R=L/2=20$ and $N_{grid}=500$ grid points.
By calculating the Fourier transform of $h^{\alpha\beta}(r)=g^{\alpha\beta}(r)-1$ according to the Lado method \cite{Lado}
we obtain the raw data for $\boldsymbol{h}$. As last step, we smooth the data.
The raw data for $h_k^{\alpha\beta}$ plotted as functions of $k$ are superimposed by oscillations stemming from
an interplay of uncertainties of $h^{\alpha\beta}(r)$ at large $r$ and the real space cutoff $R$.
These oscillations are most significant at small $k$. We smooth them out by taking the arithmetic average of the upper and
lower envelope functions. With this result we calculate the raw data for $\boldsymbol{S}_k$. Now, for the first few of the lowest $k$ grid points
the positive definiteness of $\boldsymbol{S}_k$ may be violated. We repair this as follows: let $l$ be the smallest index for which
$\boldsymbol{S}_{k_j}\succ\boldsymbol{0}$ holds for all $j\geq l$. We set $\boldsymbol{S}_{k_j}=\boldsymbol{S}_{k_{l+1}}$ for $j\leq l$.
The whole procedure can be fully automatized.

\section{Weak mixing limit}
\label{wml}

The evaluation of Eq.~(\ref{slope}) at $x_s=0$ and $x_b=0$ follows exactly the steps described in the Appendix of Ref.~\cite{Hajnal09}
with the following modifications: the packing fraction $\varphi$ has to be replaced by the
variable $\Gamma$.
Eqs.~(A29) and (A31) in Ref.~\cite{Hajnal09} have to be replaced by the zeroth order $T/2$-HNC closure
\begin{eqnarray}
\label{closure0}
\ln[(\boldsymbol{h}^{(0)})^{\alpha\beta}(r)+1]&=&(\boldsymbol{h}^{(0)})^{\alpha\beta}(r)
-(\boldsymbol{c}^{(0)})^{\alpha\beta}(r)\nonumber\\
&&-(\boldsymbol{u}_{eff}^{(0)})^{\alpha\beta}(r),
\end{eqnarray}
\begin{equation}
\label{closure0a}
(\boldsymbol{u}_{eff}^{(0)})^{\alpha\beta}(r)=
\frac{2}{\{\pi n^{(0)}\}^{3/2}}\frac{\chi_{\alpha}\chi_{\beta}}{\chi_b^2}\frac{\Gamma}{r^3},
\end{equation}
and the first order $T/2$-HNC closure
\begin{eqnarray}
\label{closure1}
(\boldsymbol{h}^{(1)})^{\alpha\beta}(r)&=&\{(\boldsymbol{h}^{(0)})^{\alpha\beta}(r)+1\}
\{(\boldsymbol{h}^{(1)})^{\alpha\beta}(r)\nonumber\\
&&-(\boldsymbol{c}^{(1)})^{\alpha\beta}(r)
-(\boldsymbol{u}_{eff}^{(1)})^{\alpha\beta}(r)\},
\end{eqnarray}
\begin{equation}
\label{closure1a}
(\boldsymbol{u}_{eff}^{(1)})^{\alpha\beta}(r)=2\{1-\delta\}(\boldsymbol{u}_{eff}^{(0)})^{\alpha\beta}(r).
\end{equation}
Furthermore, Eq.~(A32) in Ref.~\cite{Hajnal09} has to be replaced by $n^{(0)}=1$ and $n^{(1)}=0$,
since we have chosen $1/\sqrt n$ as unit length.




\bibliographystyle{elsarticle-num}



\end{document}